\documentclass[aps, twocolumn]{revtex4}
\usepackage[english]{babel}
\usepackage{graphicx}
\usepackage{dcolumn}
\usepackage{color}
\usepackage[utf8]{inputenc}
\usepackage{indentfirst}
\usepackage{amsmath, bm}

\def\beq#1{\begin{equation}\label{#1}}
\def\eeq{\end{equation}}
\begin{document}
\title{Spectral caustic in two-color high harmonic generation: role of Coulomb effects}

\author{V A Birulia$^{1}$, V V Strelkov$^{2,1}$}
\address{$^1$Moscow Institute of Physics and Technology (State University), Dolgoprudny, 141700 Moscow Region, Russia}
\address{$^2$Prokhorov General Physics Institute of the Russian Academy of Sciences, 119991 Moscow, Russia}
\begin{abstract}\noindent 
The high-order harmonic spectrum generated in two-color intense laser field under certain conditions has a pronounced maximum caused by the so called spectral caustic. Using a numerical solution of the 3D time-dependent Schr\"odinger equation we study the width of the maximum and the degree of the generation enhancement due to the caustic as a function of the fundamental intensity, wavelength, and atomic ionization potential. It is shown that the degree of this enhancement can be well quantitatively characterized by a single parameter: the ratio of the radius of the free electronic oscillation in the laser field to the atomic size. This behavior can be explained by the Coulomb attraction of the photoelectron to the parent ion, whereas the effect of this attraction on the width of the maximum is negligible. Choosing the field parameters providing a pronounced enhancement, we calculate the high harmonic macroscopic response taking into account the transient phase-matching. We show that the caustic feature can be used to provide a quasi-monochromatic XUV source with an almost linear frequency vs. time dependence. Such source can be used, in particular, for the time-resolved single-shot XUV imaging.    
\end{abstract}
\maketitle
\noindent

\section*{Introduction}

The phenomenon of high-order harmonic generation (HHG) via interaction of intense laser fields with gases and plasma gains a number of interesting features when two generating fields of different frequencies are used~\cite{old1, old2, old3}. Using the two-color driving field allows achieving an increase of the generation efficiency due to both microscopic and macroscopic aspects~\cite{theory, molecule, Platonenko99, Kim, wave-mixing0, enhancement, propagation, supercontinuum, caustic, two-color_plasma, Strelkov2016, Quasi-phase-matching},
 fine control of the electronic quantum trajectories~\cite{symmetry, holography}, spatial separation of the processes involving different number of quanta~\cite{wave-mixing}. Moreover, the HH spectrum generated in the two-color field contains more spectral components. In particular, using the fundamental and its second harmonic to generate HH one obtains not only odd, but also even harmonics of the fundamental. This allows fulfilling the resonant conditions for HHG in certain cases~\cite{res1,res2,res3}. In the temporal domain adding the second harmonic to the generating field changes the periodicity of the attosecond pulse emission from the fundamental half-cycle to the full cycle~\cite{atto_two-color1,atto_two-color2, atto_two-color3}; this facilitates generating a single attosecond pulse~\cite{dog1, dog2, dog3, two-color_ph-m, atto_two-color4, atto_two-color5}. Moreover, using two circularly polarized driving fields rotating in opposite directions provides a unique possibility of generating circularly-polarized harmonics~\cite{circ1}. 

A pronounced maximum in the high-frequency part of the harmonic spectrum was observed~\cite{caustic} for a certain ratio of the generating fields’ intensities and phase difference between the fundamental and the second harmonic.  This maximum is attributed within the three-step HHG model~\cite{simple-man} to the properties of the electronic trajectories in the continuum: under the used generating fields’ configuration {\it many} electronic trajectories come back to the origin with the same kinetic energy, leading to increased generation efficiency for a narrow group of harmonics. More strictly,  the kinetic energy of the returning electrons as a function of the return time (or as a function of the detachment time) has a {\it flat-top} maximum (namely in this maximum the second derivative is zero). Thus the origin of the generation efficiency enhancement can be understood as the ‘spectral caustic’ ~\cite{caustic}. The intensity of the narrow group of the harmonics enhanced due to this phenomenon can be an order of magnitude~\cite{caustic} (or even more~\cite{atto_Xe}) higher than the average intensity outside this spectral region.  This effect has been used recently in the HHG spectroscopy~\cite{caustic1,caustic2}.    

The role of the Coulomb effects in HHG is actively studied both theoretically and experimentally~\cite{Goreslavski Paulus, Popruzhenko Bauer, Lai, Xie, coulomb1, coulomb2, coulomb3, coulomb4, Platonenko_delay, Strelkov_delay}. However, these studies are mostly focused at the case of a one-color generating field.  

In this paper we study the phenomenon of the spectral caustic in the HHG using a numerical solution of the 3D time-dependent Schrodinger equation (TDSE) for an atom in the external field(s). We compare classical and TDSE predictions concerning the driving fields' parameters leading to the enhanced HHG efficiency for the highest harmonics. Studying the degree of this enhancement, we find that it is very sensitive to the Coulomb attraction, and that the role of the Coulomb effects is not completely characterized with the Keldysh parameter. Another parameter is suggested to describe this role. Finally, considering the macroscopic aspects of the problem, we show that the quasi-monochromatic XUV source with a time-dependent frequency can be designed using the caustic feature and discuss the perspectives of such source in the imaging applications of the HHG. 

\section{Caustic spectral feature for different values of the Keldysh parameter}

We consider HHG in a two-color field which is the sum of the fundamental field and its second harmonic: 
\beq{}
E(t)=E_0(t)[cos(\omega_0 t)+a\cdot cos(2\omega_0 t + \varphi)], 
\label{(4)}
\eeq
 where $a$ is the amplitude ratio of the second harmonic and the fundamental field,  $\varphi$ is the relative phase, $\omega_0$ is the frequency of the fundamental field, and the amplitude of the field $E_0(t)$ is given by eq.~(\ref{pulse}) in Appendix. The amplitude has sine-squared leading and trailing edges (with duration $\tau_{edge}$), and is constant within the central part of the pulse during time interval $\tau_{top}$.

\begin{figure}
\begin{minipage}[h]{0.48\linewidth}
\center{\includegraphics[width=1.0\linewidth]{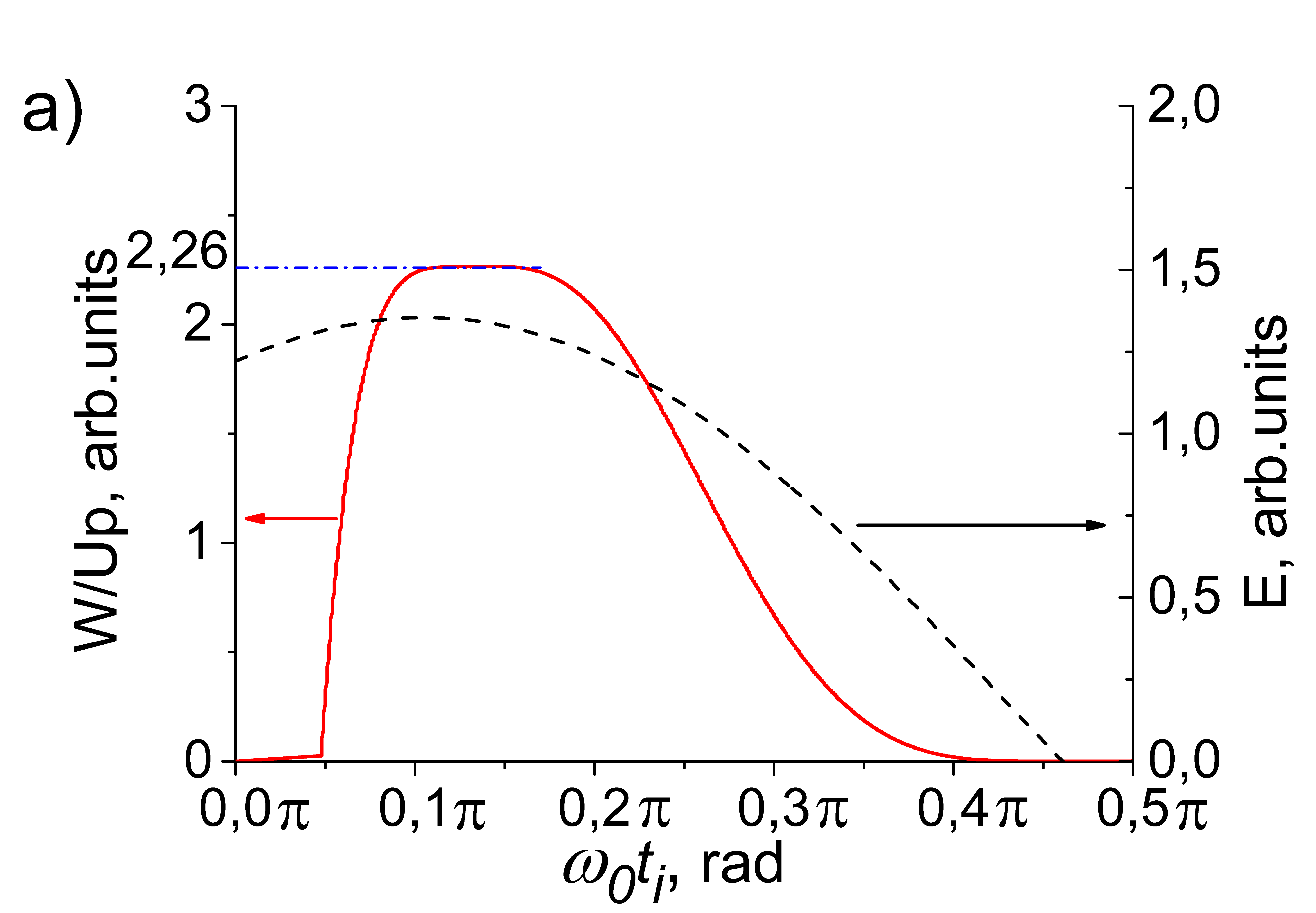}} \\
\end{minipage}
\hfill
\begin{minipage}[h]{0.48\linewidth}
\center{\includegraphics[width=1.0\linewidth]{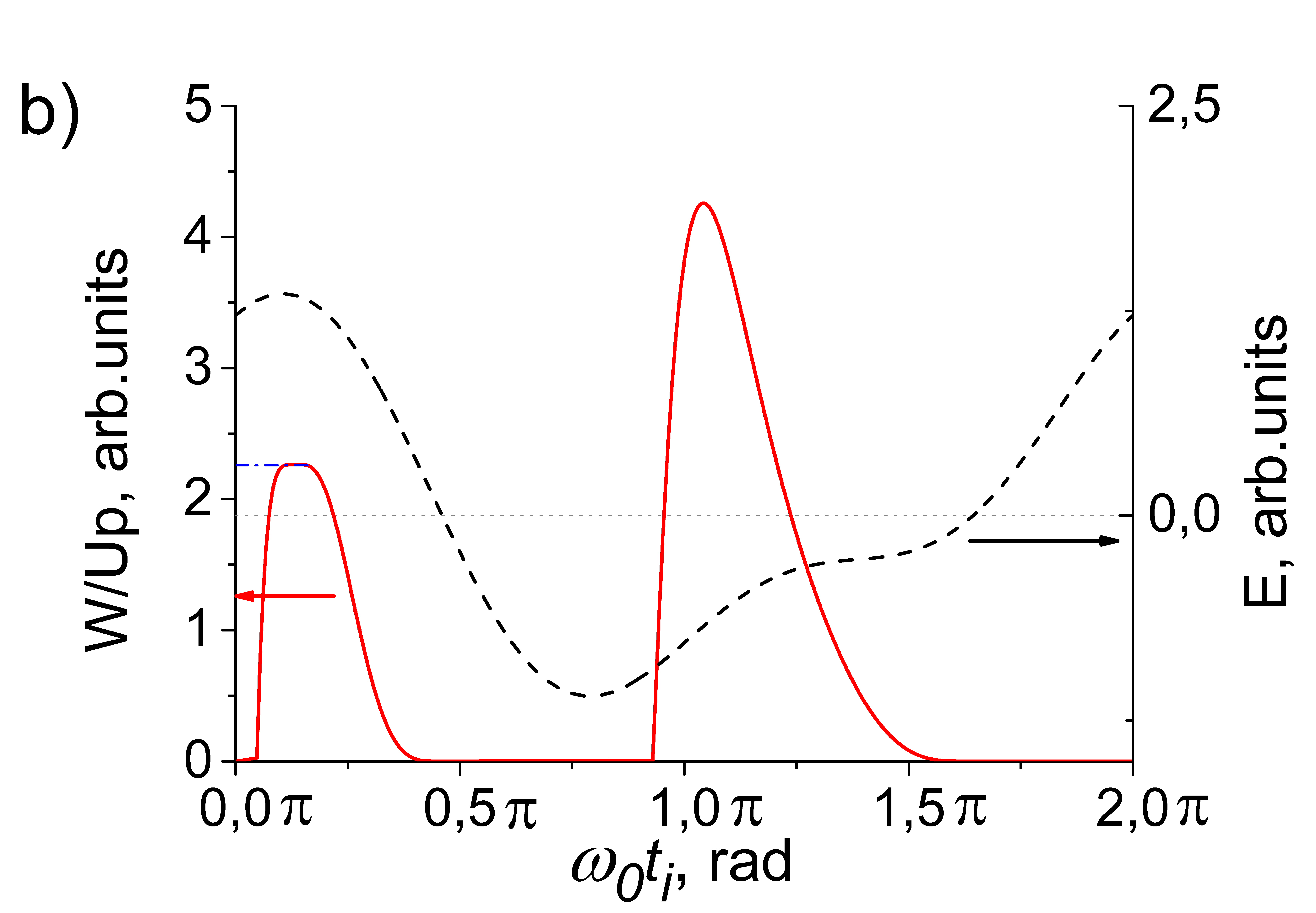}} \\
\end{minipage}
\vfill
\begin{minipage}[h]{0.48\linewidth}
\center{\includegraphics[width=1.0\linewidth]{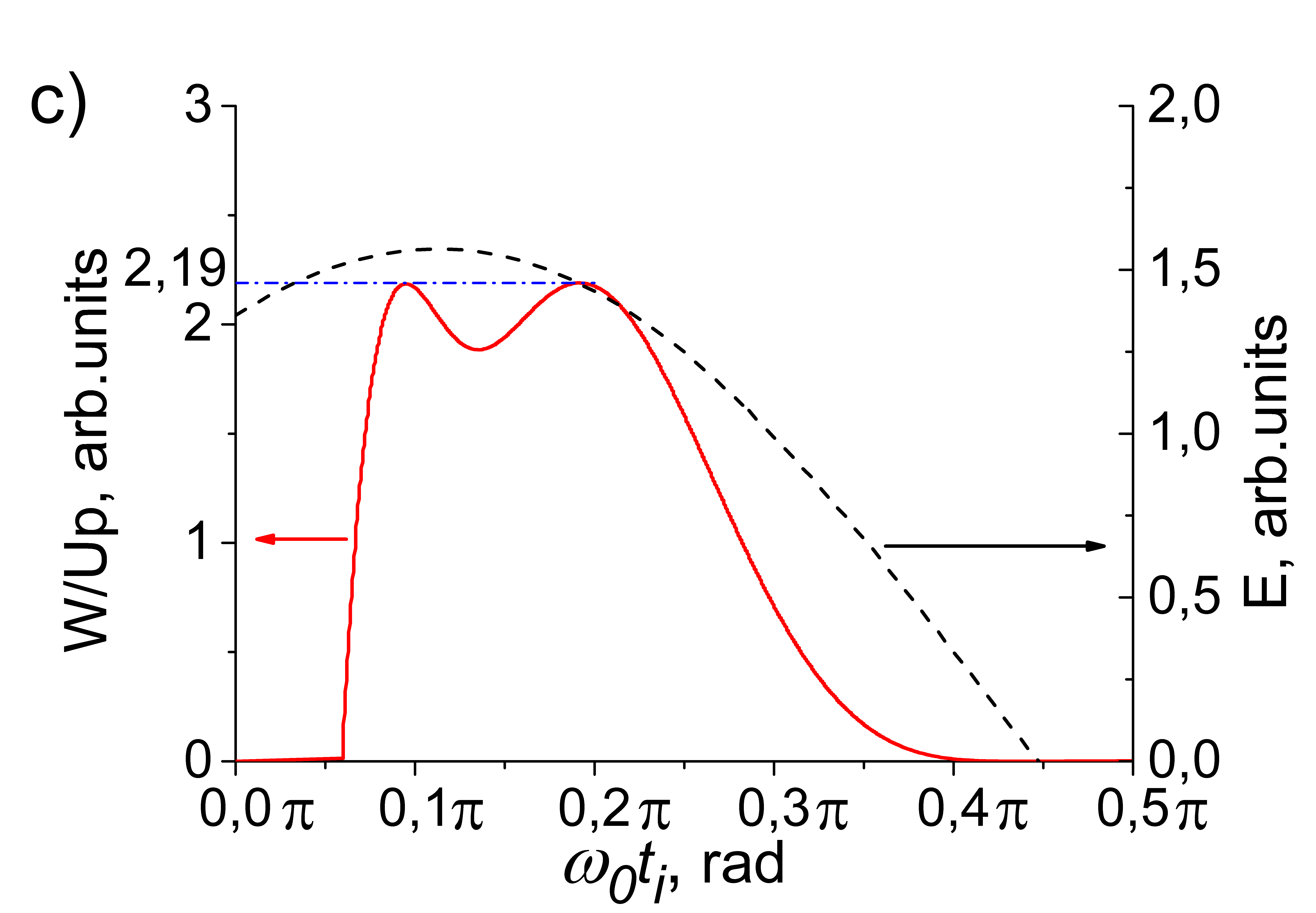}} \\
\end{minipage}
\hfill
\begin{minipage}[h]{0.48\linewidth}
\center{\includegraphics[width=1.0\linewidth]{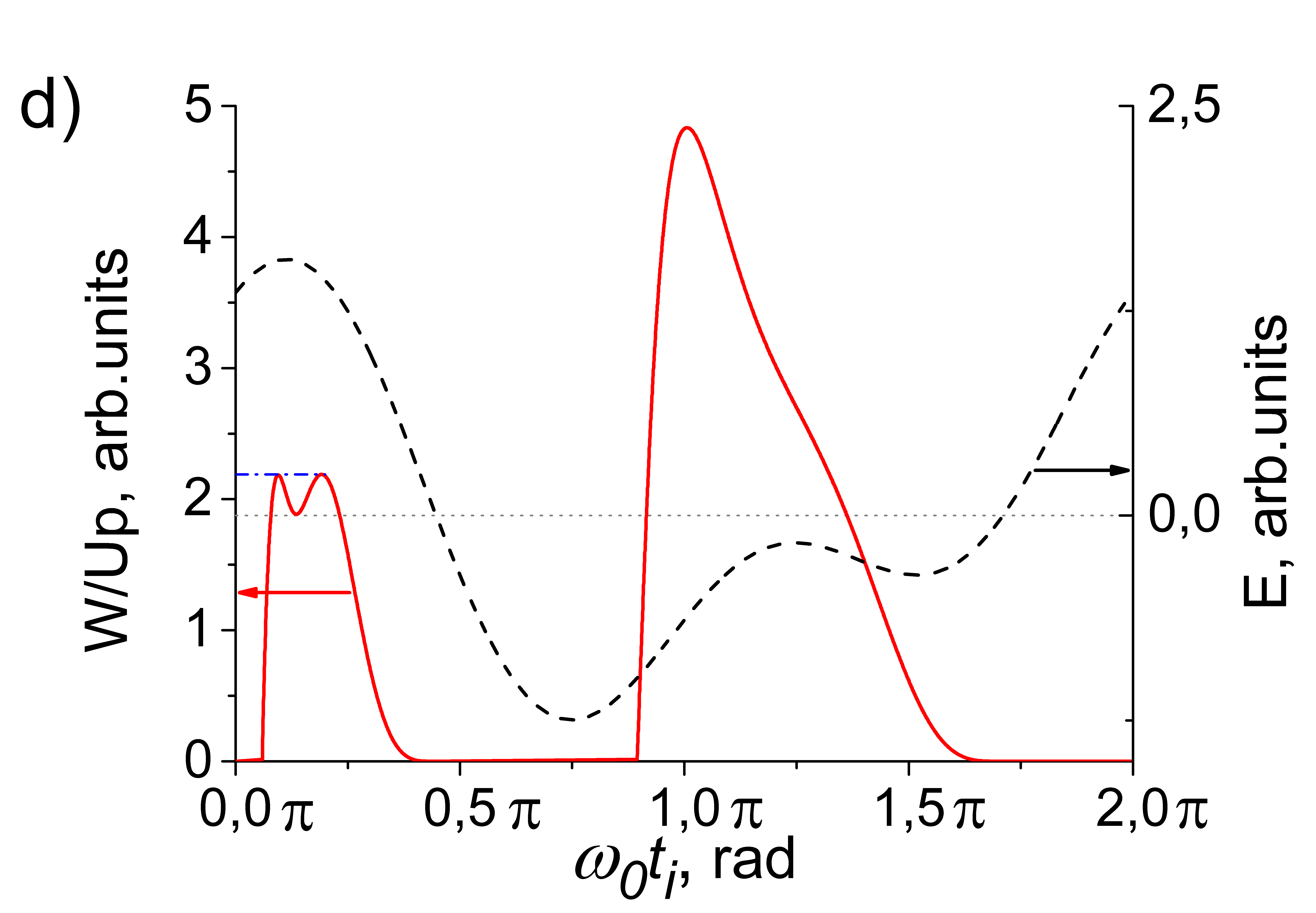}} \\
\end{minipage}
\caption{ \textit {The generating two-color field $E$ (black dashed line) and the normalized kinetic energy of the recombining electron $W$ (red continuous line) as a function of the ionization time $t_i$. The generating field is given by the expression~(\ref{(4)}) with: $a=0.44,  \varphi=-1.04$ (a, b) and $a=0.65,  \varphi=-0.98$ (c, d). The figures in the left column (a, c) and in the right column (b, d) differ only in the range of $\omega_0 t_i$ shown.}}
\label{fig:WandE}
\end{figure}

Fig.~\ref{fig:WandE} shows the results of the electronic motion study in the field~(\ref{(4)}) within the framework of the simple-man approach~\cite{simple-man}:the electron has zero velocity and zero coordinate immediately after the ionization,  the influence of the parent ion’s Coulomb attraction on the electron after ionization is neglected, the returning electron emits a photon with energy which is 
the sum of the ionization energy $I_p$ and the kinetic energy $W$ acquired by the electron during its motion in the continuum. In this figure $W$ is normalized to the ponderomotive energy of the fundamental radiation $U_p=E_0^2/4\omega_0^2$ (atomic units are used throughout the paper). As discussed in \cite{caustic}, for $a=0.44$ and $\varphi=-1.04$ the kinetic energy of the recombining electron is practically constant for a certain interval of ionization time values, leading to enhanced generation efficiency for the group of harmonics with the photon energy of about $I_p+2.3\cdot U_p$. Note that the instantaneous strength of the electric field also has its maximum within this time interval, so the ionization probability for such $t_i$ is high.  Fig.\ref{fig:WandE} (b, d) shows that harmonics with a significantly higher energy are generated for $\omega_0 t_i\approx 1.1\pi$. However, the generation of such harmonics isn't efficient, because the field strength at the instant of ionization is low, and so is the ionization probability.

 Fig.~\ref{fig:WandE} (c , d) shows the results found for parameters $a$, $\varphi$ leading to two close maxima in the kinetic energy profile. The parameters are chosen so that the kinetic energies at the maxima are the same. As we shall see below, for such parameters the caustic-like maximum in the HHG spectrum of the highest harmonics is also present. This effect is observed because within a quantum-mechanical picture the dependencies presented in Fig.~\ref{fig:WandE} a) and c) are hardly distinguishable due to the uncertainty principle (within a certain range of the parameters $a$ and $\varphi$).  

\begin{figure}
 \center{\includegraphics[width=0.7\linewidth]{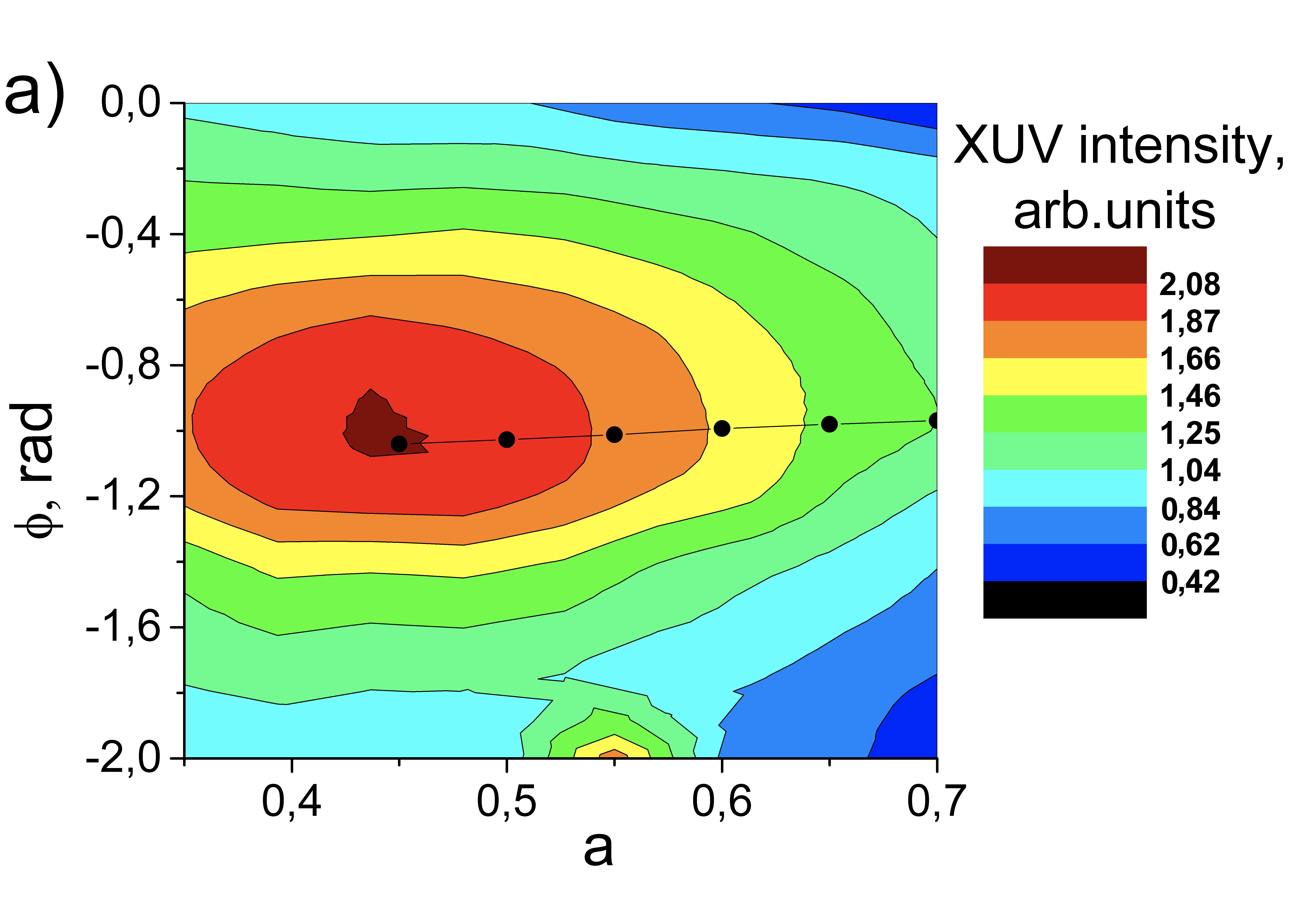}}
 \center{\includegraphics[width=0.7\linewidth]{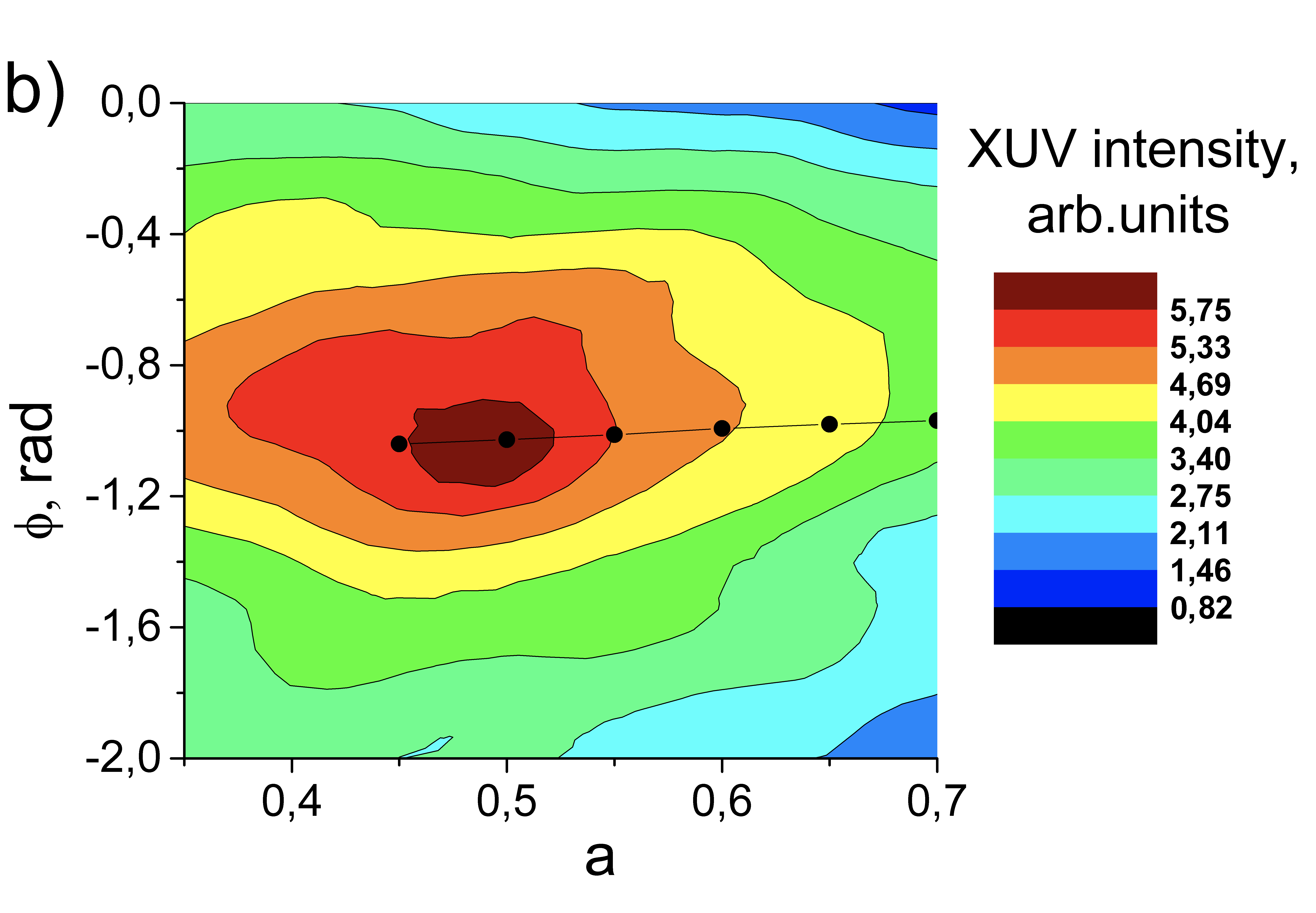}}
 \center{\includegraphics[width=0.7\linewidth]{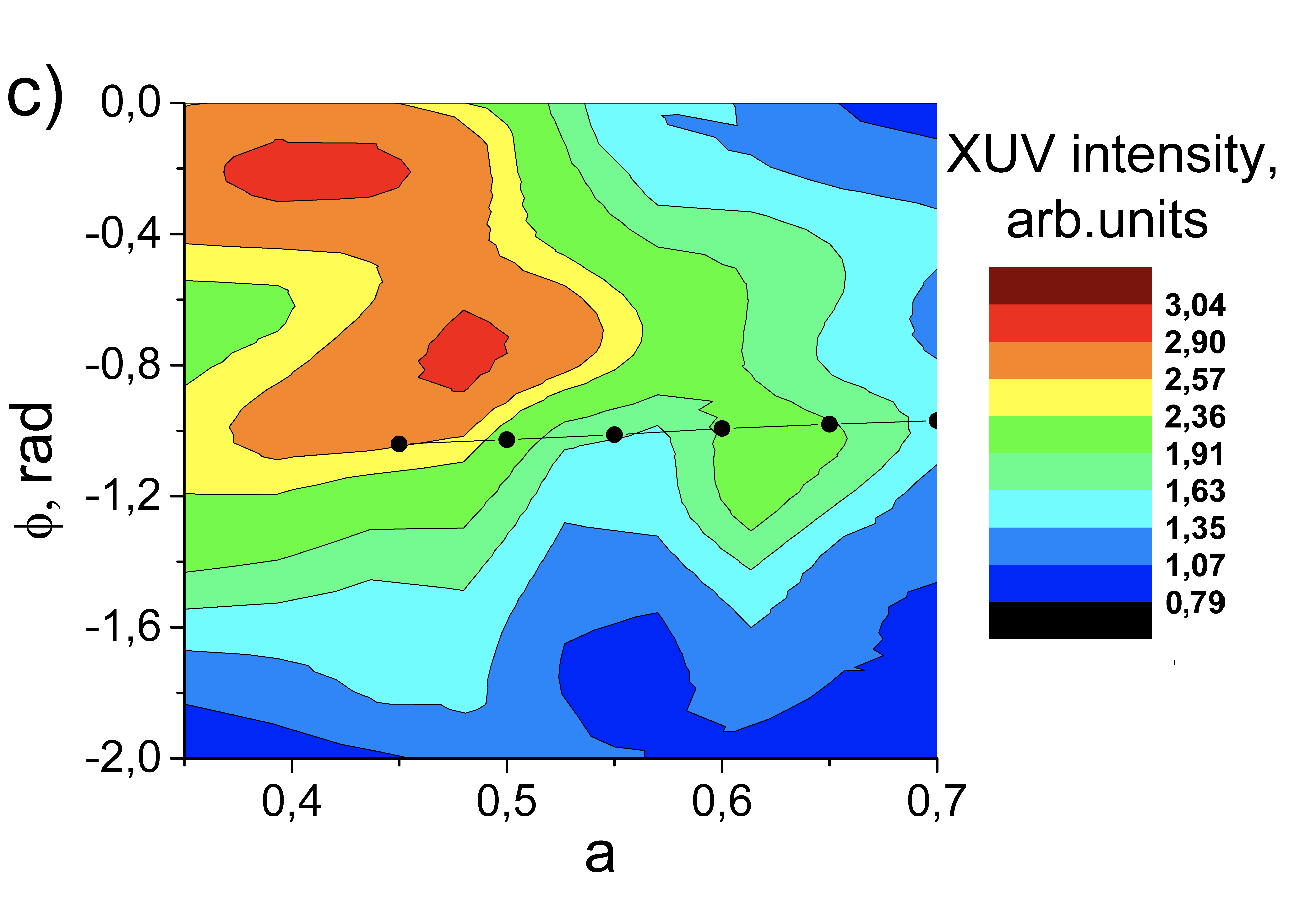}}
\caption{The total intensity of the highest harmonics calculated numerically via 3D TDSE solution for xenon using eq.~(\ref{(5)}) for different parameters of the two-color generating field $a$ and $\varphi$. The line shows the  parameters for which the two maxima of the classical kinetic energy of the returning electron have the same value (see Fig.~\ref{fig:WandE} and text for more details). The fundamental intensity is $0.5\times10^{14}$ W$\cdot$cm$^{-2}$, the fundamental wavelength is $\lambda=1.83$ $\mu m$  which corresponds to the Keldysh parameter $\gamma=0.62$ (a), $\lambda=1.52$ $\mu m$ ($\gamma=0.75$) (b), $\lambda=1.30$ $\mu m$ ($\gamma=0.87$) (c).}
\label{fig:maingraph}
\end{figure}

Our further studies are based on the numerical solution of the 3D TDSE for a model atom in the field given by eq.~(\ref{(4)}) with $\tau_{edge}=3 T_0$ and $\tau_{top}= 8 T_0$, where $T_0$ is the fundamental cycle. The details of the numerical approach are given in the Appendix (section~\ref{Num}).

First, we study the HHG enhancement due to the spectral caustic for different values of the Keldysh parameter~\cite{Keldysh} $\gamma=\sqrt{\frac{I_p}{2 U_p}}$. To characterize the enhancement we calculate the total energy of the highest-order harmonics:  
\beq{(5)}
W_{q1,q2}=\sum \limits_{q=q_1}^{q_2} {\cal I}_q 
\eeq
where ${\cal I}_q$ is the intensity of the $q$-th harmonic, $q_1=[(I_p+U_p)/\omega_0]$, $q_2=[(I_p+2.5 U_p)/\omega_0]$ (here $[$~$]$ means the nearest integer). The lower limit of the range is chosen so as to discard the low harmonics, which are not due to the tunnel ionization, and the upper one – so as to encompass the maximum energy of the harmonics which are enhanced due to the use of the two-color field, namely approximately $ I_p+2.3\cdot U_p$ (see Fig.\ref{fig:WandE}). The total energy is shown in Fig.~\ref{fig:maingraph} for different parameters of the two-color field. Note that the range from $q_1$ to $q_2$ corresponds to different harmonic orders for each wavelength of the generating radiation. Namely, it corresponds to harmonics 37 - 72 for $\lambda = 1.83$ $\mu m$ in graph (a), to harmonics 28 - 47 for $\lambda=1.52$ $\mu m$ in graph (b), and to harmonics 23-35 for $\lambda=1.30$ $\mu m$ in graph (c).  

To compare the classical and the quantum mechanical results we also show in Fig.~\ref{fig:maingraph} the parameters of the two-color field for which the kinetic energy vs. ionization time diagram has maxima at the same level (see Fig.\ref{fig:WandE} (a,c)). 

We can see a good agreement between the classical and the quantum mechanical calculations for the low Keldysh parameters (graphs (a) and (b)). Namely, the maximum energy of the highest harmonics is achieved for the parameters close to  $a=0.44$ and $\varphi = -1.04$, i.e. for the values which are used in Fig.\ref{fig:WandE} (a), where the diagram of the kinetic energy vs. ionization time has a practically flat top. 

 However, for Fig.~\ref{fig:maingraph} (c) the correspondence with the classical calculation disappears because for the Keldysh parameter close to unity the ionization does not occur in the tunneling regime, and the classical approach discussed above is not applicable.

\section{Degree of the HHG enhancement due to the spectral caustic}

In this section we study the role of the Coulomb effects on the spectral caustic, focusing at the conditions corresponding to the tunneling ionization regime. As we have shown in the previous section, in this regime the prediction of the classical calculations concerning the parameters $a$ and $\varphi$ maximizing the energy of the highest harmonics agrees with the numerical results. So below throughout the paper we are using these values of the parameters.   

\begin{figure}
\begin{minipage}{0.8\linewidth}
\center{\includegraphics[width=1.0\linewidth]{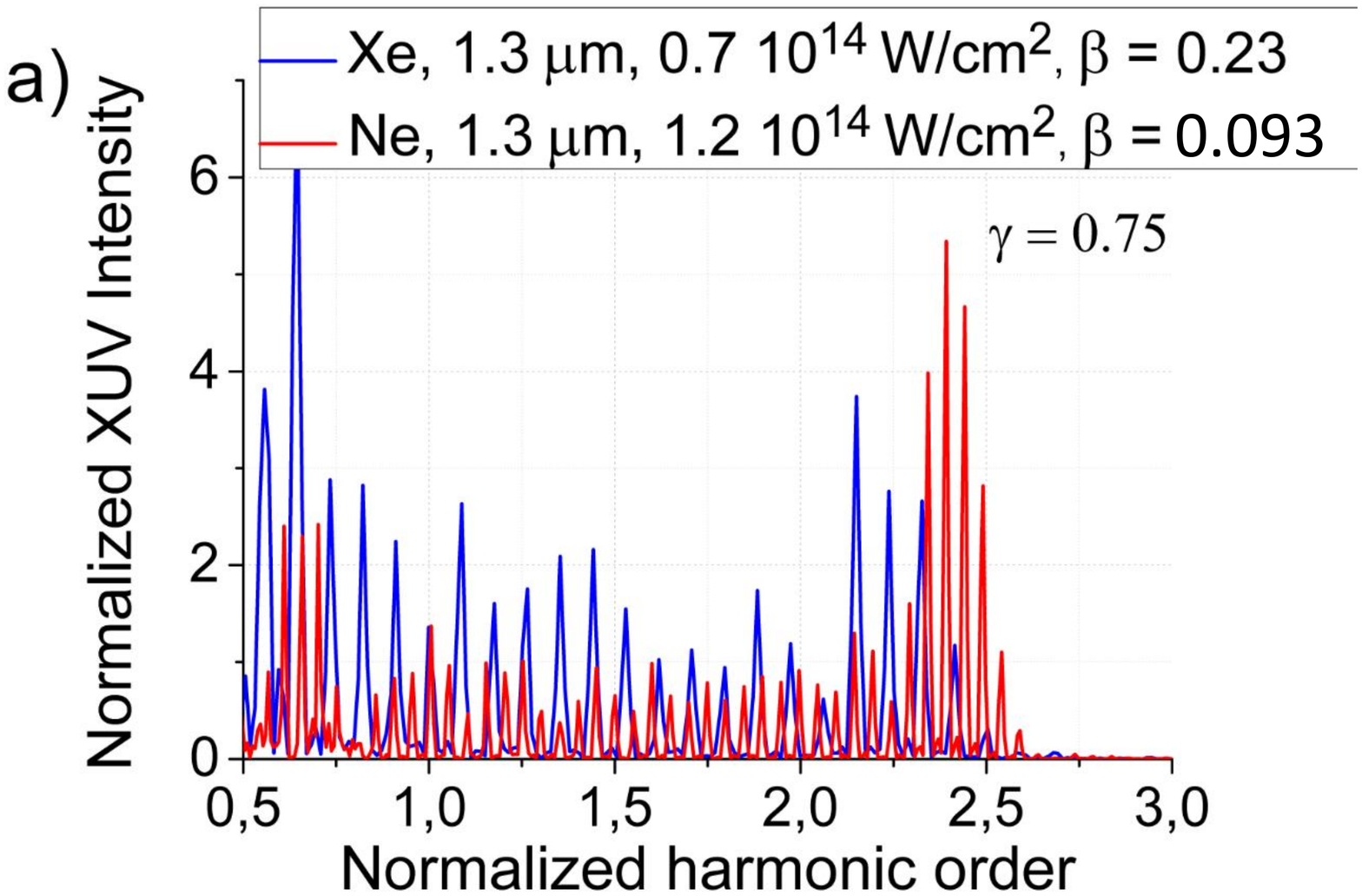}}
\end{minipage}
\hfill
\begin{minipage}{0.8\linewidth} 
\center{\includegraphics[width=1.0\linewidth]{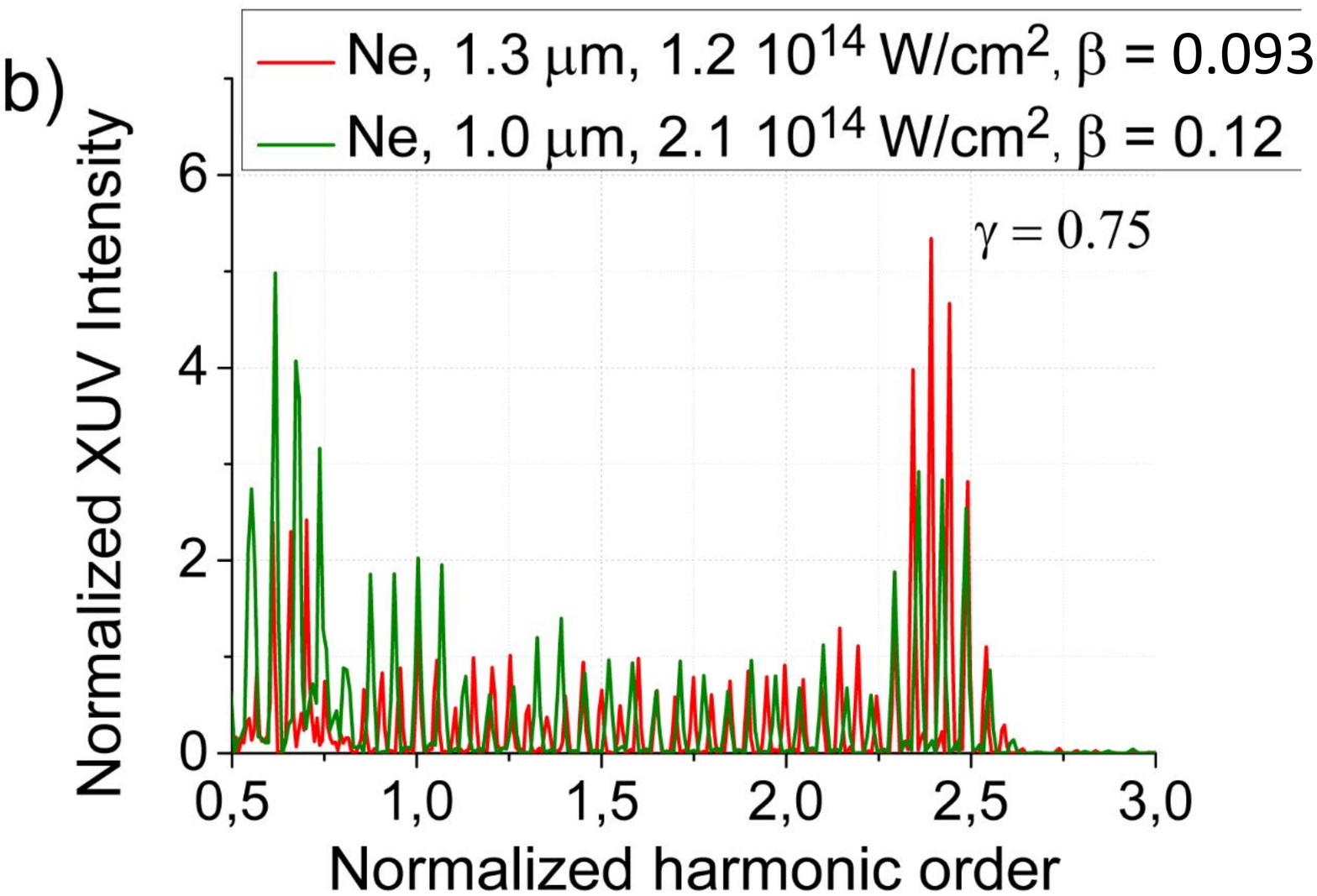}}
\end{minipage}
\caption{The normalized (see text for more details) spectra calculated via 3D TDSE numerical solution for neon and xenon in the two-color field with parameters $a=0.44, \varphi=-1.04$, and the Keldysh parameter $\gamma=0.75$. The spectra for the same wavelength and different intensities and atomic species (a) and for neon at two different wavelengths and intensities (b).}
\label{fig:spektrs}
\end{figure}

Fig.\ref{fig:spektrs} shows the HHG spectra calculated for xenon and neon for the different  fundamental field intensities and wavelengths chosen so that the Keldysh parameter is the same. To compare the widely differing spectra we normalize the axes as following: $(\omega-I_p)/U_p$ is plotted along the horizontal axis, where $\omega$ is XUV frequency,  and ${\cal I}(\omega)/{\cal I}_{av}$ is plotted along the vertical one, where  ${\cal I}(\omega)$ is XUV intensity and 
\beq{I_av}
{\cal I}_{av}= W_{q1,q2}/(q_2-q_1+1).
\eeq
For this normalization we use the spectral range corresponding to the highest plateau harmonics but excluding those near the spectral caustic feature:   $q_1=[(I_p+1.5 U_p)/\omega_0]$ and $q_2=[(I_p+2.1 U_p)/\omega_0]$. 

We can see in  Fig.\ref{fig:spektrs} that due to the spectral caustic feature a group of harmonics with photon energy close to $ I_p+2.3\cdot U_p$ is generated more efficiently than the other high harmonics, but the degree of this enhancement depends on the fundamental intensity, wavelength, and the generating atom. We discuss below how to characterize these dependences quantitatively.

Let $x_0$ is the coordinate of the electron’s exit from under the barrier created by the Coulomb and laser fields. Then $x_0\approx I_p/E_0$. The Coulomb field strength in $x_0$ is $E_{Col}= 1/x_0^2=E_0^2/I_p^2$. We introduce the parameter $\beta$ as the ratio of this strength to that of the laser field:
\beq{(6)}
\beta=\frac{E_{Col}}{E_0}=\frac{E_0}{I_p^2} 
\eeq

Note that in the expression for $\beta$ the intensity of the laser field is in the numerator, which might seem counterintuitive. This means that the stronger is the laser field, the greater is the effect of the ion at the exit point of the electron from beneath the barrier. This is explained by the potential barrier becoming narrower with increasing light intensity, so that the electron exits into the continuum at a point closer to the core.

In the discussion below we shall use the normalized field~\cite{Perelomov}: $\mathcal{E}=E_0/\sqrt{2\cdot I_p^3}$ . It allows to compare the processes in atoms with different ionization potentials. The parameter $\beta$ is expressed in terms of the normalized field as follows: 
\beq{(7)}
\beta=\frac{\mathcal{E}\cdot 2 \sqrt{2}}{\sqrt{I_p}} 
\eeq
Ionization rates of different atoms for the same values of $\mathcal{E}$ are close to each other. So if we compare different atoms in a field with the same normalized amplitude, then the parameter $\beta$ is lower for higher $I_p$.

Fig.\ref{fig:spektrs} (a) shows the HHG spectra for xenon and neon for the same fundamental wavelength and Keldysh parameter and different intensities so that $\beta$ is higher for xenon. We can see that the caustic feature for neon is more pronounced. The spectra in Fig.\ref{fig:spektrs} (b) are calculated for neon for different intensities and wavelengths (corresponding to the same Keldysh parameter). Again, the caustic feature is more pronounced for the lower $\beta$. Thus we conclude that in case of identical $\gamma$ the parameter $\beta$ can be used for estimating qualitatively the influence of Coulomb attraction on HHG: for smaller $\beta$ the role of the Coulomb field is smaller.

\begin{figure}
\center{\includegraphics[width=0.9\linewidth]{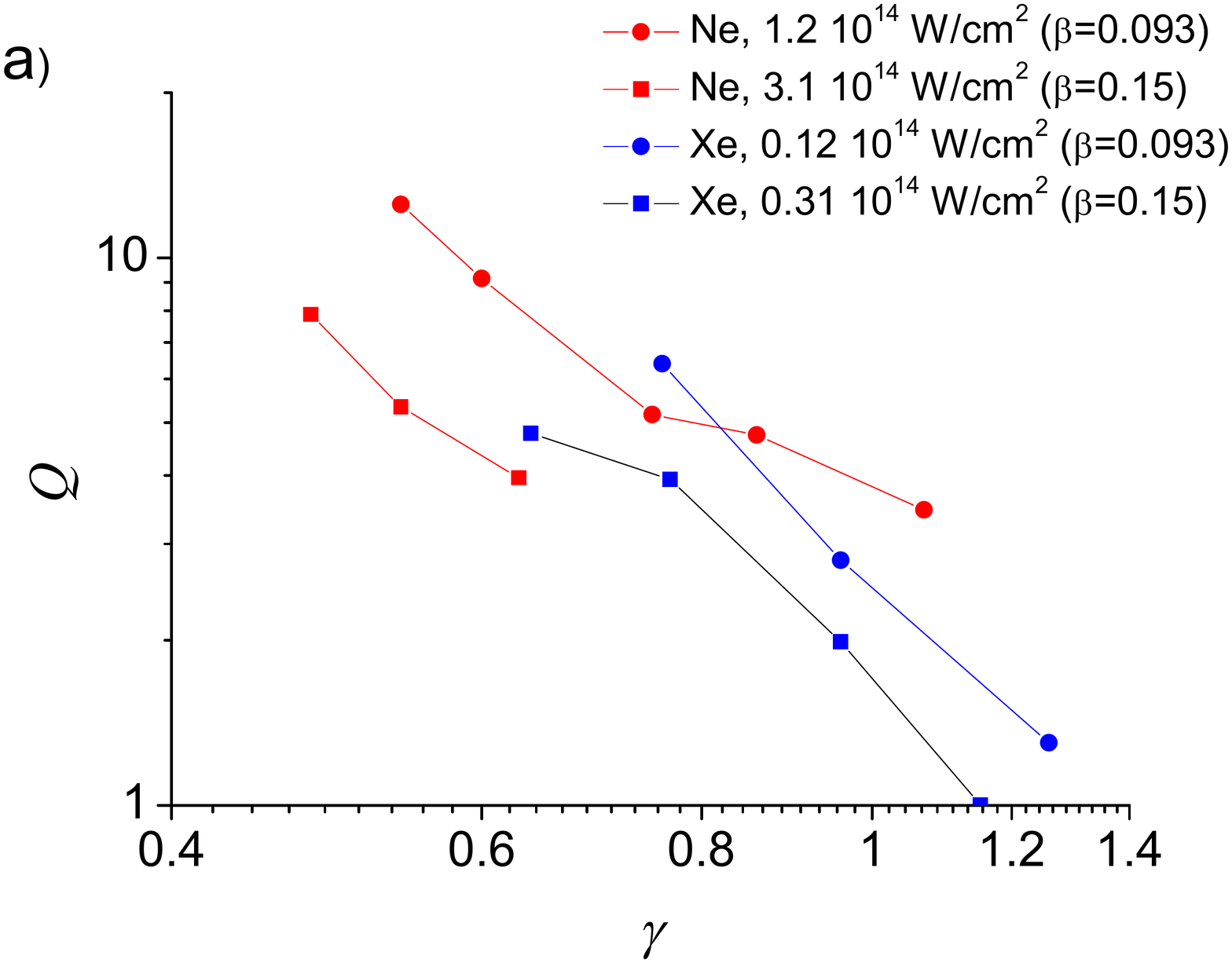}}
\center{\includegraphics[width=0.9\linewidth]{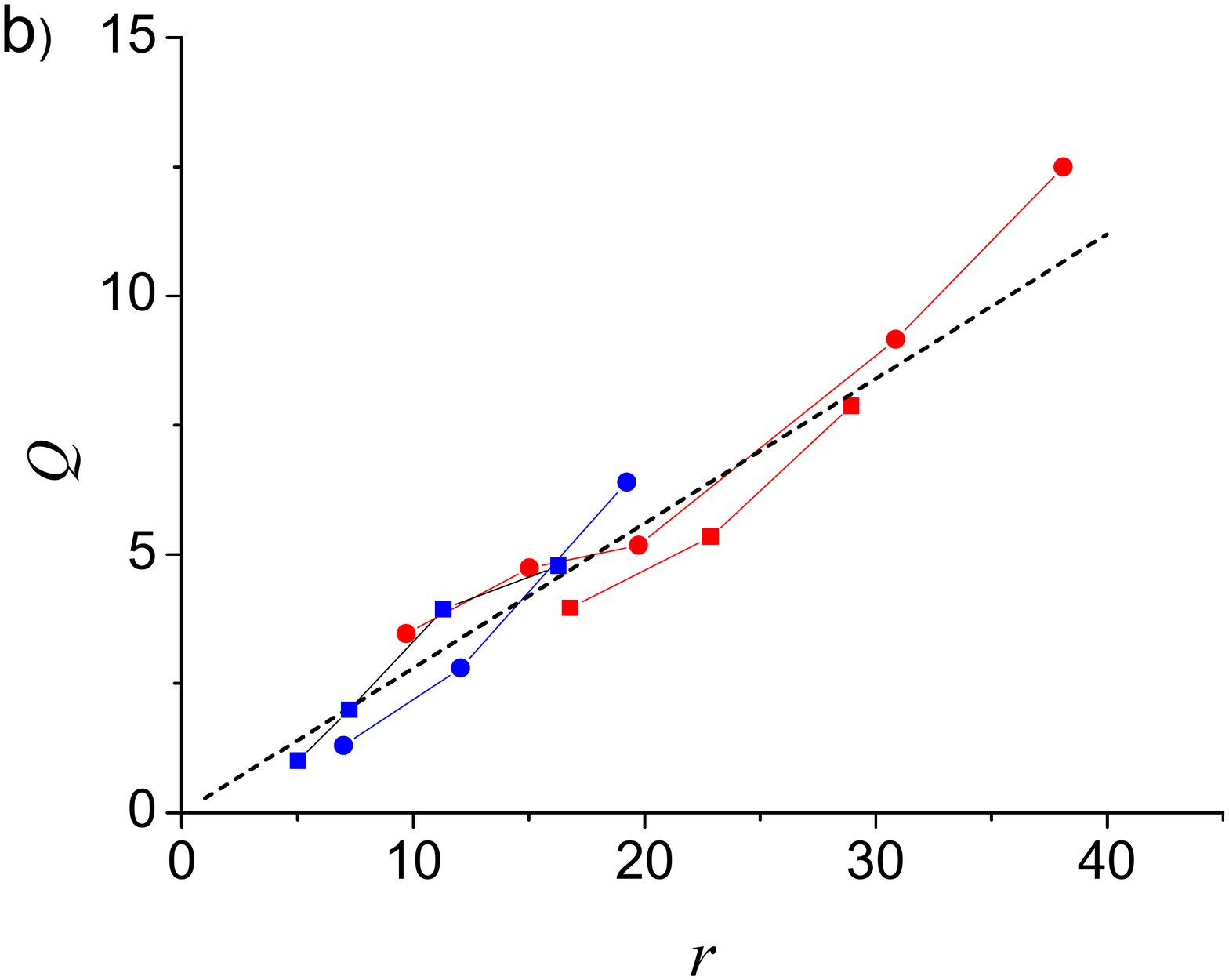}}
\caption{The degree of the HHG enhancement due to the spectral caustic (given by the expression~(\ref{Q})) calculated for neon and xenon as a function of the Keldysh parameter (a) and the parameter $r$ given by eq.~(\ref{r}) (b). Fundamental intensities and corresponding values of the $\beta$ parameter are presented in the graph. Dotted line in graph (b) shows approximation of this value by eq.~(\ref{Q_approx}).}
\label{fig:ratio}
\end{figure}

Now we shall study the behavior of the caustic maximum for different Keldysh parameters. Namely, we calculate the ratio of the maximum harmonic intensity in the vicinity of the caustic feature ${\cal I}_{max}$ to the average intensity of the plateau harmonics given by eq.~(\ref{I_av}):
\beq{Q}
Q={\cal I}_{max}/{\cal I}_{av}.
\eeq
Fig.~\ref{fig:ratio} shows this ratio calculated via 3D TDSE numerical solution for neon and xenon under different fundamental wavelength and intensities (corresponding to different $\beta$ parameters). The dependence is approximated by the following equation:
\beq{Q_approx}
Q\approx C r
\eeq
here $C=0.28$ is a constant, 
\beq{r}
r=\frac{1}{\beta \gamma^2}=\frac{E_0 I_p}{2 \omega_0^2}=\frac{R_{osc}}{2 R_{at}}
\eeq
where $R_{osc}=E_0/\omega_0^2$ is the radius of the free oscillations of the electron in the field (the excursion length) and $R_{at}=1/I_p$ is the ''atomic radius'' (this is the distance from the origin where the Coulomb potential is equal to the ionization potential). 

We can see that eq.~(\ref{Q_approx}),~(\ref{r}) approximate the numerical results very well. The parameter $r$ has a straightforward physical sense: it characterizes the ratio of the distance where the electronic motion is mainly controlled by the laser field to that where it is controlled by the Coulomb field.  Thus one can conclude that the HHG enhancement due to the caustic feature is limited by the Coulomb effects, at least in the (rather wide) range of the parameters used in our calculations. 

Such an important role of the Coulomb attraction in this problem can be understood as following. After tunneling the electron is accelerated by the laser field, but the Coulomb attraction suppresses this acceleration to some extent. For a one-color field the role of the Coulomb attraction in HHG was studied in many papers, as we mentioned in the Introduction. In particular, it was shown~\cite{Platonenko_delay} that the Coulomb attraction leads to an effective delay before the start of the really free motion of the electron, or, in other words,  it leads to the classical trajectories being populated by the electrons which tunneled slightly earlier than the classical "detachment time". For a one-color field this enhances the contribution of the so called short electronic trajectory to HHG~\cite{Strelkov_delay}.      

In the two-color field the kinetic energy profile of the returning electron has a flat top, but then goes down to zero even steeper than in the one-color field, see Figs.~\ref{fig:WandE}. This defines the dramatic role of the Coulomb effects in the two-color case. If the "delay before the free motion" is rater pronounced, then the mostly populated electronic trajectories do not correspond to this flat-top region any more, so the maximum due to the caustic feature is smeared. This disappearence of maximum in the spectrum for high $\beta$ can be seen in Fig~\ref{fig:spektrs}.

Note that the parameter given by eq.~(\ref{r}) is widely used, in particular, to quantitatively explain the simple-man model~\cite{simple-man}. However, the applicability of this parameter to achieve a good quantitative agreement with the numerical results is remarkable.   

\section{Spectral width of the caustic-induced maximum}

\begin{figure}
\center{\includegraphics[width=0.9\linewidth]{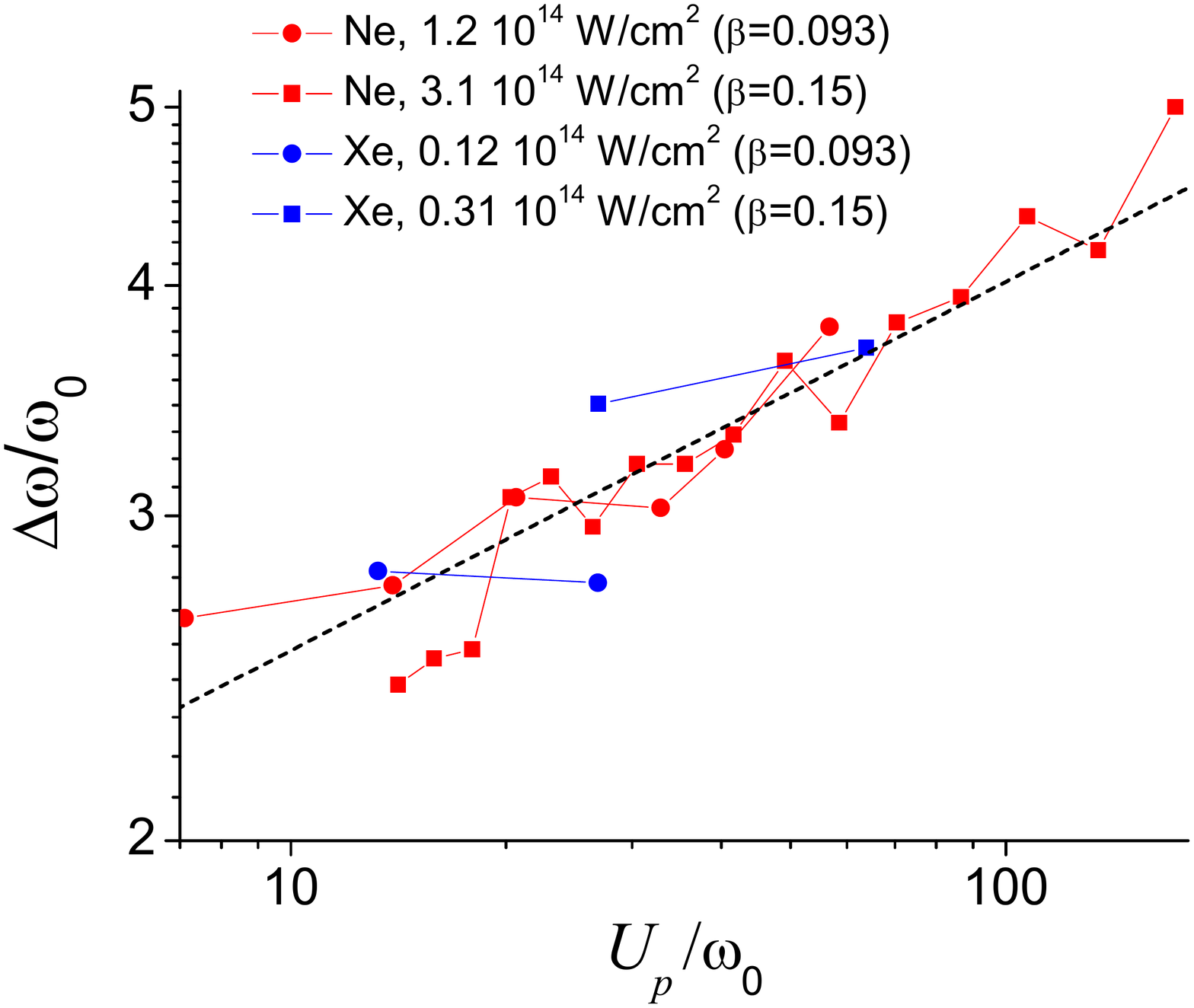}}
\caption{The normalized width of the maximum due to the spectral caustic for neon and xenon as a function of the parameter $U_p/\omega_0$. Fundamental intensities are the same as in Fig.~\ref{fig:ratio}. Dotted line shows the approximation of the width by eq.~(\ref{width2}).}
\label{fig:width}
\end{figure}

As we can see in Fig.~\ref{fig:spektrs} the maximum due to the spectral caustic includes several harmonics. To study the width of the maximum in more details we calculate its FWHM for different generation conditions. The results are shown in Fig.~\ref{fig:width}. We can see that the normalized width does not change much within a wide range of the conditions.

To study this behavior analytically we consider the emission of the XUV with an instantaneous frequency $I_p+W(t)$ calculated within the simple-man model.  We assume that the spectral width of the signal emitted during a time interval $\Delta t$ near $W(t)$ maximum is:
\beq{width1}
\Delta \omega^2=\frac{a_1^2}{\Delta t^2} +(a_2 \Delta t^4)^2
\eeq
where the first term corresponds to the uncertainty principle and the second one describes the spectral width due to the instantaneous frequency variation within the interval $\Delta t$:

\beq{}
a_2=\frac{1}{4!} \frac{\partial^4}{\partial t^4} W(t)
\eeq
 
Function $W(t)$ can be written as $W(t)=U_p w(\omega_0 t)$ (see Fig.~\ref{fig:WandE}) so 
\beq{d4W}
\frac{\partial^4}{\partial t^4} W(t) = U_p \omega_0^4 \frac{\partial^4 w (\tau)}{\partial \tau^4}
\eeq
where the last term does not depend on $U_p$ and $\omega_0$.

Assuming that the instantaneous XUV intensity within the time interval $\Delta t$ is approximately constant we conclude that the maximal spectral intensity is achieved when the spectral width given by equation~(\ref{width1}) is minimal, thus under $\frac{\partial}{\partial \Delta t} \left(\Delta \omega^2 \right)=0$. From this condition we find relation between $a_1$ and $a_2$ using eq.~(\ref{d4W}), and finally find that $$\Delta \omega \propto \sqrt[5]{U_p \omega_0^4}.$$ So
\beq{width2}
\Delta \omega/ \omega_0 = B \sqrt[5]{U_p/\omega_0} 
\eeq
where $B$ is a constant. In Fig.~\ref{fig:width} we present this approximation with $B=1.6$. We can see that the approximation agrees with the numerical results.  

Note that the expression~(\ref{width2}) was found without taking the Coulomb effects into account. Good agreement with the numerical results shows that, whereas the maximum's {\it intensity} is affected by the Coulomb attraction, the latter does not in fact influence the {\it width} of the maximum. One can attribute this to the fact that the attraction mainly modifies the distribution of the electronic density between the different electronic trajectories in the continuum, but it has almost no impact on the kinetic energy of the returning electron as a function of the return time (given by the function $W(t)$ within the simple-man approach).

The ponderomotive energy is much higher than the fundamental photon energy in typical HHG conditions. Thus eq.~(\ref{width2}) shows that (at least for the case of linearly polarized generating fields) the maximum due to the spectral caustic includes several harmonics and it can hardly consist of a single harmonic. This is important, for instance, for the spectroscopic applications of this phenomenon.

\section{Quasi-monochromatic XUV source with time-varying frequency}

In the previous sections we have found the requirements for the laser field and the generating atom parameters allowing to obtain a pronounced maximum in the HH spectrum due to the spectral caustic. Such pronounced maximum can be utilized in several applications~\cite{caustic1, caustic2}. In this section we study the perspective of using such HHG conditions to provide a quasi-monochromatic XUV source with time-varying frequency.

\begin{figure}[h!]
\begin{minipage}[h]{0.9\linewidth}
\center{\includegraphics[width=1.0\linewidth]{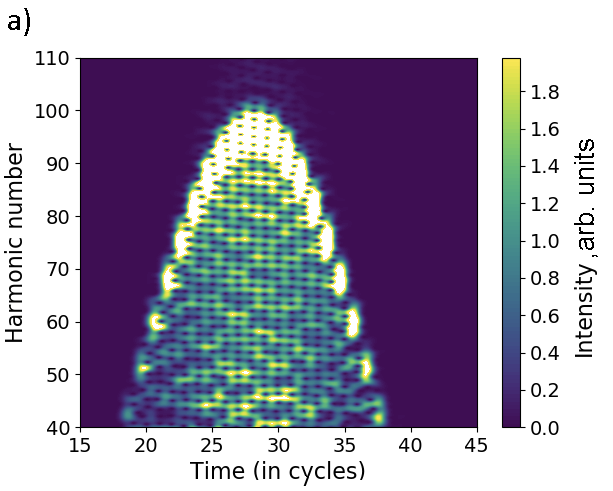}}
\end{minipage}
\hfill
\begin{minipage}[h!]{0.9\linewidth}
\center{\includegraphics[width=1.0\linewidth]{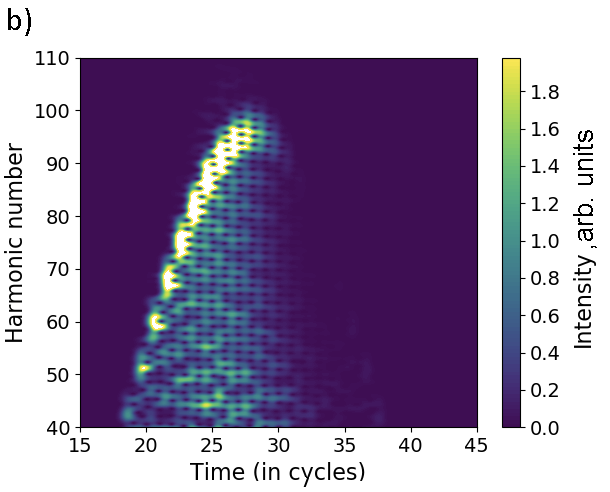}}
\end{minipage}
\caption{Gabor transform of the microscopic high-frequency response calculated for neon in the two-color field (a); Gabor transform of the macroscopic response calculated taking into account the transient phase-matching (b). The peak fundamental intensity is $2.0\times10^{14}$ W/cm$^{2}$, the wavelength is 1.3~$\mu$m, the full pulse duration $2 \tau_{edge}$ is 56 fundamental cycles (duration at the half of the peak intensity level is 90 fs), the gas density is $N = 5.0\times 10^{17}$ cm$^{-3}$ and the target thickness is $z_{full} = 0.1$ cm.}
\label{fig:gabor}
\end{figure}

\begin{figure}[h!]
\begin{minipage}[h]{0.9\linewidth}
\center{\includegraphics[width=1.0\linewidth]{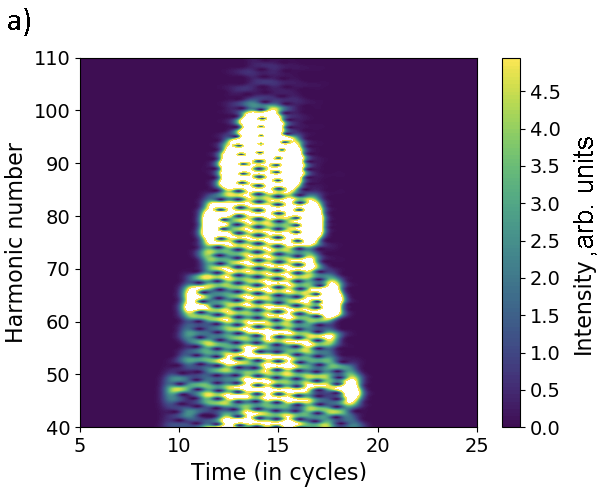}}
\end{minipage}
\hfill
\begin{minipage}[h!]{0.9\linewidth}
\center{\includegraphics[width=1.0\linewidth]{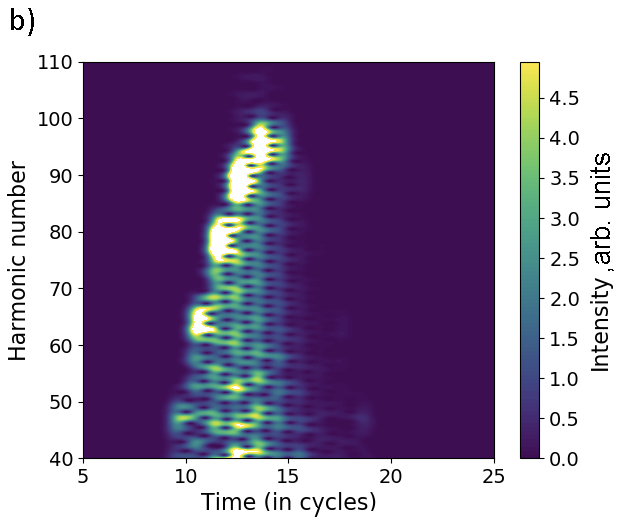}}
\end{minipage}
\caption{The same as on Fig.\ref{fig:gabor} for the two times shorter pulse and the two times higher gas density.}
\label{fig:gabor2}
\end{figure}

Up to now we were using a flat-top laser pulse, providing (at least for certain conditions discussed above) a pronounced maximum in the XUV spectrum due to the caustic feature. Using a more realistic bell-shape laser pulse leads to less pronounced maximum, because the position of the maximum in the spectrum varies in time due to the variation of the laser intensity in time. This dependence makes the maximum in the total spectrum less distinct. However, this dependence can be utilized to build an XUV source with a time-varying frequency. In this section, using the above-discussed conditions for HHG providing the pronounced maximum, we shall study the transient frequency of such source.  Moreover, we shall show that taking into account transient phase-matching in the macroscopic XUV response one can provide an almost linear variation of this frequency, and discuss the perspectives of such source in HHG imaging.  

Using the bell-shape fundamental pulse $E_0(t)$ (given by eq.~(\ref{pulse}) with $\tau_{top}=0$) we calculate the microscopic response $\mu(t)$ and find its Gabor transform $S(\omega, t)$ (the duration of the temporal window in the Gabor transform is one fundamental cycle). Its intensity $|S(\omega, t)|^2$ is shown in Fig.~\ref{fig:gabor} (a) and Fig.~\ref{fig:gabor2} (a). We can see that the position of the maximum in the spectrum varies with time. In the XUV macroscopic response the transient phase-matching changes the XUV intensities. Namely, at the front of the laser pulse the ionization degree is low, and the phase-matching is good; as the ionization takes place the phase-matching becomes worse and the macroscopic signal decreases~\cite{Kazamias, two-color_ph-m}.  In particular, the gas target parameters (to be more precise, product of the density and length) can be chosen so that the generation is phase-matched only at the front edge of the pulse and suppressed near the maximum and at the falling edge. Our approach for the transient phase-matching calculation is described in Appendix (section~\ref{ph-m}). At Fig.~\ref{fig:gabor} (b) and Fig.~\ref{fig:gabor2} (b) we show the Gabor transform of the macroscopic response $|S(\omega, t)_{prop}|^2$ calculated with eq.~(\ref{eq:sinc}) from the Appendix. 

We can see that for the used conditions the macroscopic signal is quasi-monochromatic, and its frequency varies in time almost linearly. Changing the experimental parameters (laser pulse duration and intensity, gas pressure and thickness of the target) one can control the frequency variation range and rate. Such XUV source can have various applications. One of them is the XUV time-resolved imaging.   The direct link between the XUV frequency and time would allow single-shot imaging with femtosecond temporal resolution. Namely,  an image obtained using a certain spectral range corresponds to a certain time instant; making several images in different spectral ranges one can use them as a serial frames describing the dynamics of the imaged system.  

\section{Conclusion}

Using the numerical 3D TDSE solution we have studied the properties of the high-order harmonics generated by the fundamental laser field and its second harmonic. The maximum in the highest part of the HHG spectrum caused by the spectral caustic is investigated. We have found the parameters of the two-color field (namely, the ratio of the field strengths and the phase difference between the fields) which ensure this maximum. These parameters calculated assuming classical free electronic motion agree very well with the numerical TDSE solution result when the Keldysh parameter is much less than unity. It is shown that the maximum typically consists of several harmonics and that its width can be well understood within the simple-man approach and thus it is not affected much by the Coulomb effects. The expression we have found for the width of the maximum agrees well with numerical results.

However, even if the Keldysh parameter is less than unity (i.e. under the tunneling ionization conditions) the degree of the generation enhancement due to the spectral caustic depends very much on the ionization potential of the atom, on the fundamental frequency and wavelength. This can be explained by the influence of the Coulomb attraction on the electronic motion in the continuum: when the Coulomb effects are negligible many electronic trajectories come back to the origin with the same kinetic energy producing the spectral caustic; when these effects are non-negligible, the kinetic energies become different, the maximum is smeared and thus much less pronounced. We introduce the parameter $r$ (which is the ratio of the radius of the free electronic oscillation in the laser field to the atomic size) and show that this parameter describes quantitatively the enhancement of the generation efficiency due to the spectral caustic. This allows us finding the conditions where the maximum due to the spectral caustic is very pronounced. Under such conditions the enhanced harmonics can be used, in particular,  to provide a quasi-monochromatic XUV source. Taking into account the temporal variation of the fundamental intensity as well as the transient phase-matching we have found the conditions when the instantaneous XUV frequency depends on time almost linearly. Such source can be used, in particular, for the time-resolved single-shot XUV imaging.    

\section*{Acknowledgments}
The numerical calculations of the microscopic XUV response were funded by the Russian Science Foundation (grant No 16-12-10279). The studies of the macroscopic signal for the XUV imaging applications were supported by RFBR (grant No 16-02-00858).

\section*{Appendix}

\subsection{Temporal envelope of the laser pulse}

The slow-varying amplitude of the laser field in eq.~(\ref{(4)}) is given by the following equation: 

\beq{pulse}
\begin{array}{l}
E_0(t)= \\
  \begin{cases}
  A\sin^2\left(\frac{\pi t}{2 \tau_{edge}}\right), & 0 \leq t \leq \tau_{edge} 
  \\
 A , & \tau_{edge} <  t \leq \tau_{edge}+\tau_{top} 
  \\
 A\sin^2\left(\frac{\pi (t - 2\tau_{edge} - \tau_{top})}{2 \tau_{edge}}\right), & \tau_{edge}+\tau_{top} < t \leq 2 \tau_{edge}+\tau_{top}  
\end{cases}
\end{array}
\eeq
where $A$ is the amplitude of the field, $\tau_{edge}$ is the duration of leading/trailing edge of the pulse, $\tau_{top}$ is the time interval, when $E_0$ remains constant. If $\tau_{top}$ is zero, then the pulse is bell-shaped.

\subsection{Calculation of the microscopic response}
\label{Num}

To calculate the quantum-mechanical microscopic response we solve the 3D single-active electron TDSE for a model atom in a linearly polarized laser field: 
\beq{TDSE}
i \dot{\psi} (r,z,t)=\left(\frac{\hat p^2}{2}+V(r,z)+E(t)z\right)\psi(r,z,t)
\eeq
where $r$ is the polar radius, $z$ is the coordinate along the direction of the electric field,  $\hat p$ is the momentum operator. 

For xenon we use the soft Coulomb potential: $$V(r,z)=-\frac{1}{\sqrt{b^2+z^2+r^2}}$$ with $b = 0.14$ which allows reproducing the actual ionization potential for this atom.  

For neon a more complicated potential is used: $$ V(r,z)=-\frac{1+\exp{(-\frac{r}{d}})}{\sqrt{b^2+z^2+r^2}}$$ with parameters $b = 0.5, d = 2.7$. These parameters allow reproducing the neon ionization potential. 

The microscopic response is given by the quantum mechanical average of the dipole moment. According to the Ehrenfest theorem its second derivative is equal to the average of the force acting on the electron:
\begin {equation*}
\ddot{\bm{\mu}}(t) = \mathbf{f}_{full}(t),
\end{equation*}
where
\beq{}
\mathbf{f}_{full}(t) = -\langle \psi(\mathbf{ r}, t)\lvert \mathbf{E}(t)+\frac{\mathbf{ r}}{r^3}\rvert \psi({\bf r}, t)\rangle.
\eeq
However, in the electric dipole approximation $\mathbf{E}$ is independent of the coordinate, so the first term equals $\mathbf{E}(t)$. To find the nonlinear microscopic response we calculate:
\beq{}
\mathbf{f}(t) = -\langle \psi \lvert \frac{\mathbf{r}}{r^3}\rvert \psi \rangle .
\eeq
The method for the numerical solution of eq.~(\ref{TDSE}) is described in~\cite{JPB_num}.

\subsection{Calculation of the macroscopic response}
\label{ph-m}

The transient HHG phase-matching was considered in several papers~\cite{Constant99, Strelkov2002,Kazamias, two-color_ph-m, UFN}. Here we consider the 1D propagation. The generation of the XUV with the frequency $\omega$ is characterized by the phase mismatch  

\begin{equation}
\label{eq:wavevector}
 \delta k =  k_{\omega} - k_{\omega}^{pol},
 \end{equation}
where $k_{\omega} = n_{\omega} \omega/c$ - is the XUV wave-vector,  $n_{\omega}$ is the refraction index for this frequency, $c$ is light velocity and the wave-vector of the polarization $k_{\omega}^{pol}$ is 
\begin{equation}
 k_{\omega}^{pol} = \frac{\partial}{\partial z} \left[ \frac{\omega}{\omega_0}(k_0 z + \varphi_{las}(z, \xi)) + \varphi_{at}(z, \xi) \right]  
 \label{eq:kqpol}
\end{equation} 
where 
\begin{equation}
\xi = t - z/v_g, 
\label{eq:timexi}
\end{equation} 
$v_g$ is the group velocity, $\varphi_{las}$ is the phase of the laser field, $\varphi_{at}$ is the so called "atomic phase"~\cite{gradient, Pascal}, $z$ is the propagation direction.

We assume that the shape of the laser pulse does not change during the propagation. Then $\varphi_{las} = \omega_0 \Delta n_{nl}(\xi)  z/c$, where $\Delta n_{nl}(\xi)$ is a nonlinear change of the refraction index for the fundamental. This change is mainly defined by the ionization of the gas, so $\Delta n_{nl}(\xi) = \sqrt{1-4\pi N_e(\xi)/\omega_0^2}$, where $N_e(\xi)$ is the electronic density. 

The atomic phase depends only on the laser intensity~\cite{Pascal, Strelkov2016} (and does not depend on its phase): $\varphi_{at}(z, \xi)=\varphi_{at}(I(z, \xi))$. Assuming that the intensity does not depend on $z$ (i.e. that the generation takes place in the laser focus) we have: $\frac{\partial}{\partial z}\varphi_{at}(z, \xi) =0$.

Thus we have from eq.~(\ref{eq:wavevector}),~(\ref{eq:kqpol})  
$$
    |\delta k| = \frac{\omega}{c}|(n_{\omega}-n_0-\Delta n_{nl})|
$$

Under the ionization degree of several percents the linear gaseous dispersion $n_{\omega}-n_0$ is compensated by the plasma term $\Delta n_{nl}$. Taking into account that we deal with higher ionization degree we neglect the linear dispersion:  
\begin{equation}
\label{eq:deltak}
    |\delta k(\omega, \xi)| =2\pi \omega N_e(\xi)/(\omega_0^2\cdot c)
\end{equation}

Moreover, we neglect the XUV absorption because it is less important than the phase mismatch for the relatively high harmonics considered here~\cite{Constant99}. 

Under the used approximations the phase-matching integral leads to the phase-matching factor which is formally the same as in the stationary case, but the factor itself depends on $\xi$. Thus, one can use the Gabor transform $S(\omega, \xi)$ of the microscopic response $\mu(\xi)$ and write the macroscopic response after propagation as:
\begin{equation}
|S_{prop}(\omega, \xi)|^2= |S(\omega, \xi)|^2  \left| \frac{\sin(\delta k(\omega, \xi)\cdot z_{full}/2)}{\delta k(\omega, \xi)\cdot z_{full}/2} \right|^2 N^2
\label{eq:sinc}
\end{equation} 
where $N$ is the atomic density and $z_{full}$ is the target thickness.

Note that one more simplification that we have done is that the phase-matching is not influenced by second harmonic propagation. More accurate description assumes that the process of a $q-th$ harmonic generation in the two-color field can be described as a sum of the processes involving $q_1$ photons from the fundamental and $q_2$ photons from the second harmonic field so that $q=q_1+2 q_2$~\cite{Platonenko99,Quasi-phase-matching}. Every process is described by its detuning from the phase-matching similar to eq.~(\ref{eq:wavevector}). However, in the considered case the intensity of the second harmonic is much less than the fundamental intensity ($a^2=0.2$), so the typical number of photons from the second harmonic field is much less than that from the fundamental one. Thus all the detunings are close to each other and close to the one given by eq.~(\ref{eq:wavevector}). Omitting the change of the second harmonic phase during the propagation we have omitted the change of the microscopic response due to the variation of the phase difference between the fundamental and the second harmonic $\varphi$. However within the coherence length this variation is low, namely, it is less than the variation of the fundamental phase which is $\pi /q$. For high harmonic orders considered here such variation of $\varphi$ does not change the microscopic signal much, see Fig.~\ref{fig:maingraph}.

\end{document}